# A Review of Disease and Development

Ruiwu Liu

Supervised by Professor Markus Brueckner

*Australian National University*

Acemoglu and Johnson (2007) put forward the unprecedented view that health improvement has no significant effect on income growth. To arrive at this conclusion, they constructed predicted mortality as an instrumental variable based on the WHO international disease interventions to analyse this problem.

We replicate the process of their research and eliminate some biases in their estimate. In addition, and more importantly, we argue that the construction of their instrumental variable contains a violation of the exclusion restriction of their instrumental variable.

This negative correlation between health improvement and income growth still lacks an accurate causal explanation, according to which the instrumental variable they constructed increases reverse causality bias instead of eliminating it.

# I. Introduction

After World War II, human health greatly improved. Specifically, the worldwide average life expectancy at birth has significantly increased, from 45.7 years in 1950 to 71.7 years in 2015 (Roser, 2015). Nevertheless, the relationship between health improvement and economic development remains unclear. It is widely believed that, in general, health improvement has a positive impact on economic growth. Many economists have carried out a great deal of research to prove this. For example, Gallup and Sachs (2001, p. 91) argued that wiping out malaria in sub-Saharan Africa could increase that continent's GDP per capita growth rate by as much as 2.6 percent a year.

However, the direct impact of health improvement on the economy is uncertain because health improvement can indeed increase the productivity of workers, thus intuitively possibly increasing the accumulated wealth of society; on the other hand, with an increase of accumulated wealth, people can obtain better medical services and live in more sanitary conditions, thereby further improving their health situation. Moreover, health improvement will lead to an increase in population, which may exaggerate its impact on the growth of social wealth. In other words, as there are some undetectable reverse causalities between health improvement and economic growth, we cannot directly analyse the impact of life expectancy on GDP per capita growth.

Economists usually want to eliminate the interference of reverse causalities using instrumental variables, but we have long been unable to find a suitable and valid variable for this problem. Acemoglu and Johnson (2007) tried to analyse this problem by constructing an ingenious instrumental variable and obtained the result that there is no positive correlation between economic growth and health improvement.

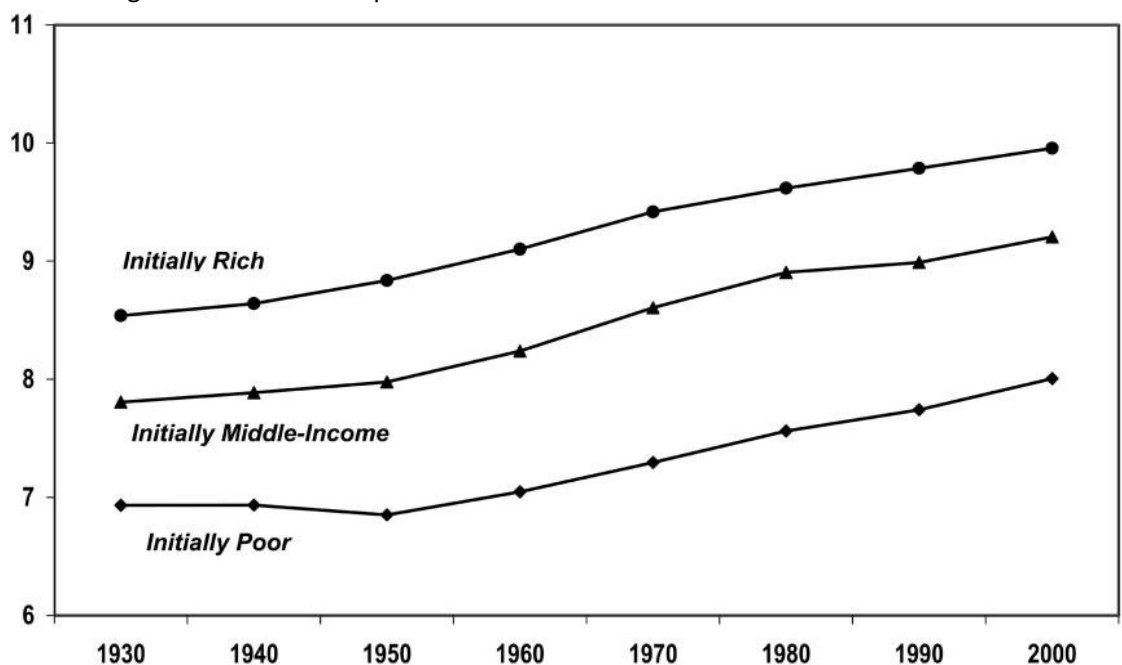

Fig. 1. Log GDP per capita for initially rich, middle-income, and poor countries.

Source: Acemoglu and Johnson (2007, p. 928)

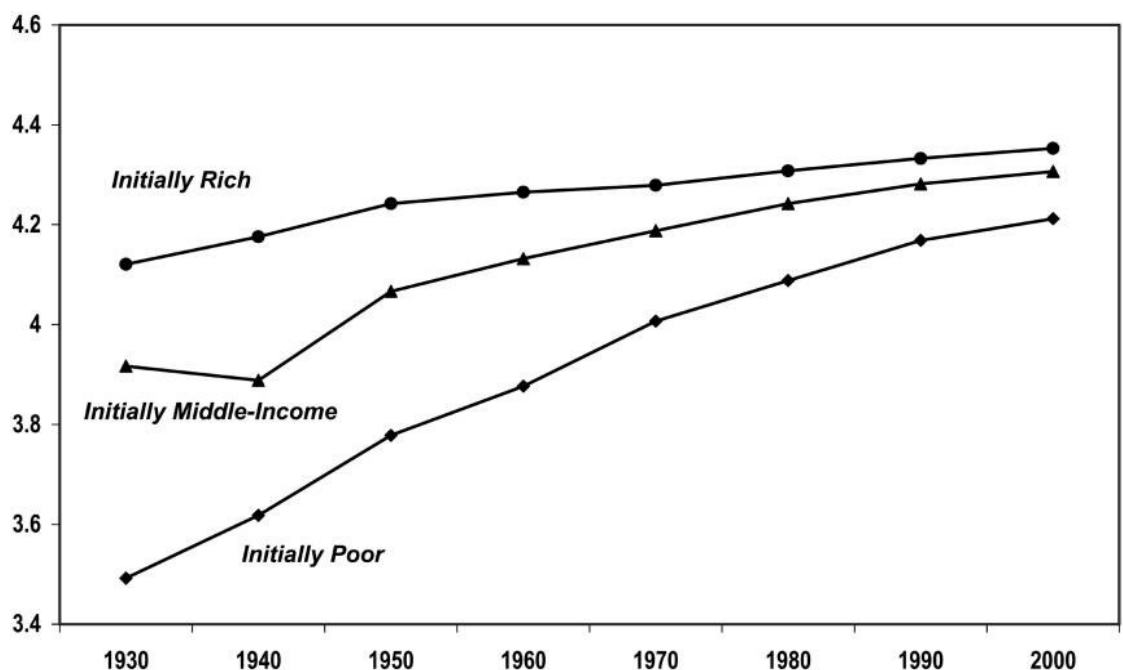
Fig. 2. Log life expectancy at birth for initially rich, middle-income, and poor countries.
Source: Acemoglu and Johnson (2007, p. 927)

Acemoglu and Johnson (2007) found that beginning in the 1940s, life expectancy converged between developed and underdeveloped economies, but the average gap in GDP per capita between rich and poor countries did not experience such a convergence (see Figures 1 and 2). That is, for low- and middle-income countries, life expectancy growth rate is relatively larger than the growth rate of GDP per capita, so that in recent years, health conditions across regions have been more balanced than economic wealth. They investigated the cause of this phenomenon and attributed it to the WHO international disease interventions started in the late 1940s. Acemoglu and Johnson (2007, pp.934-936) claimed that the WHO international disease interventions occurred for three main reasons: Firstly, there were broad innovations in medical science from the 1940s to the 1950s. Secondly, the World Health Organization was established in 1948. Thirdly, there was a change in international value. As Preston (1975, p. 243) emphasizes, "Universal values assured that health breakthroughs in any country would spread rapidly to all others where the means for implementation existed" after the 1930s. Consequently, Acemoglu and Johnson (2007, p. 925) constructed the predicted mortality instrument using preintervention mortality from assorted diseases and dates of international interventions.

We now replicate Acemoglu and Johnson (2007)'s main work in our replication section and review it with comments by Bloom, Canning and Fink (2014). In our extension section, we first analyse some of the potential biases in Acemoglu and Johnson (2007) before discussing the invalidity of the instrumental variable they constructed.

## II. Data and Estimating Framework

*A. Data*

The key variables used in our analysis include GDP, GDP per capita, life expectancy, population and total births, which are all available from UN data sources. For the construction of our instrumental variable, we collected mortality data for our sample epidemics from the WHO and UN (before the establishment of the WHO), and we adopted the times of global interventions for certain epidemics, which were used in Acemoglu and Johnson (2007). Africa and Eastern Europe are not in our dataset since there are too many missing values in our sample years (between 1940 and 1980) for Africa and because of concerns regarding the quality of the Eastern European Data. After excluding these countries, 47 countries remained in our main sample. We also provide some results after using available data from other countries, but the estimated results are consistent with the estimated results using our main sample.

*B. Main Model*

Our basic regression model is

$$y_{it} = \alpha x_{it} + \eta_i + \mathbf{Z}'_{it}\boldsymbol{\beta} + \mu_t + \varepsilon_{it} \qquad (1)$$

where $y_{it}$ is Log GDP per capita at time $t$ in country $i$; $x_{it}$ is Log life expectancy; $\eta$ is a full set of fixed effects from the basic solow-swan model; $\mu_t$ is other time-varying variables across different countries; $\mathbf{Z}'_{it}$ is all other control variables; and $\varepsilon_{it}$ is the error term. According to this, our health and GDP are influenced by many other variables across different countries. We estimate our regressions with fixed effects for our panel-data (Acemoglu and Johnson, 2007, p. 933).

*C. Construction of IV*

We now repeat the steps showing how we construct our instrumental variable following Acemoglu and Johnson (2007, p. 946).

The predicted mortality instrument is defined as:

$$M_{it}^I = \sum_{d \in D} [(1 - I_{dt})M_{dit} + I_{dt}M_{dFt}] \qquad (2)$$

where $D$ is the set of the 14 sample diseases (listed in table 1); $I_{dt}$ is a dummy variable of the international intervention for disease $d$ at time $t$ (we define it as being equal to 1 if there was an intervention for this certain disease); $M_{dit}$ is the actual mortality from disease $d$ at time $t$ in country $i$; and $M_{dFt}$ is the mortality rate from disease $d$ at time $t$ in the healthiest area in the world, which is coded as zero for all.

Next, we use a simple OLS test to estimate our first-stage:

$$x_{it} = \gamma M_{it}^I + \widetilde{\eta}_i + Z'_{it}\widetilde{\beta} + \widetilde{u}_t + \widetilde{\varepsilon}_{it} \qquad (3)$$

Equation (3) is defined similarly to equation (1); we use it to test the validity of instrument relevance condition for our predicted mortality.

As mentioned above, Acemoglu and Johnson (2007) claimed that the international disease interventions are not directly related to our dependent variable, they believe that predicted mortality satisfies our exclusion restriction because it is the interaction of initial disease mortality and interventions.

## III. Replication

### A. OLS Estimate for the Population-Related Outcomes

**Table 1**

**OLS Estimate: Dependent Variable: Log Population, Log Total Births and Percentage of Population under 20**

|  | A. Log population | | |
| --- | --- | --- | --- |
|  | Raw Sample | Main Sample | Developing Countries |
|  | 1960 and 2000 | 1940 and 1980 | 1940 and 1980 |
| Log Life Expectancy | 1.603*** | 1.623*** | 1.862*** |
|  | (0.301) | (0.191) | (0.267) |
| Number of Observations | 240 | 96 | 72 |
| Number of Countries | 120 | 48 | 36 |
| $R^2$ | 0.883 | 0.901 | 0.876 |
|  | B. Log number of births | | |
|  | 1960 and 2000 | 1940 and 1980 | 1940 and 1980 |
| Log Life Expectancy | 2.015*** | 2.351*** | 2.577*** |
|  | (0.396) | (0.272) | (0.403) |
| Number of Observations | 94 | 90 | 68 |
| Number of Countries | 48 | 45 | 34 |
| $R^2$ | 0.362 | 0.723 | 0.768 |
|  | C. Percentage of Population under Age 20 | | |
|  | 1960 and 2000 | 1940 and 1980 | 1940 and 1980 |
| Log Life Expectancy | 0.045 | 0.094*** | 0.124*** |
|  | (0.081) | (0.029) | (0.042) |
| Number of Observations | 80 | 80 | 56 |
| Number of Countries | 40 | 40 | 28 |
| $R^2$ | 0.827 | 0.215 | 0.195 |

Source: Based on the Acemoglu and Johnson (2007) sample.

Note: Robust standard errors are in parentheses.

*p<.1

**p<.05

***p<.01

Table 1 reports the population-related outcomes for our simple OLS estimates. Columns 1, 2 and 3 report estimated results for different samples. For example, column 2 in panel A shows that a 1% increase in life expectancy leads to a 1.6% increase in population size in our main sample. By contrast, this correlation is greater in developing countries, as column 3 shows. The results for total births are consistent with the population estimation as panel B shows, and improved health still shows a greater effect in developing countries; precisely, column 3 reports that an 1% increase in life expectancy at birth is associated with an 2.6% increase in total births in developing countries.

Panel C reports that this positive relationship remains when we estimate the population structure. For instance, column 3 suggests that a 1 percent increase in life expectancy at birth will result in a 1.2 increase in the percentage of the population under the age of 20. However, we lost the statistical significance of our estimate in column 1 when we focused on our raw sample, which may verify our assumption in the data description section.

## B. Zeroth-Stage

Firstly, we construct a zeroth-stage test to measure the relationship between international disease interventions and the sum of mortality from all 14 sample diseases (Acemoglu and Johnson, 2007, p. 948):

$$M_{dit} = \theta I_{dt} + \mu_t + \pi_t + \rho_i + \tau_{it} \qquad (4)$$

Controlling for other variables, we use equation (4) to estimate the relationship between the actual disease mortality and interventions.

**Table 2**
**Zeroth-Stage Estimates: Effect of Interventions on Disease Mortality**

|  | OLS | OLS with Lags |
|---|---|---|
| Intervention | -24.15*** | -24.47*** |
|  | (5.67) | (5.19) |
| Lagged Intervention |  | -18.81*** |
|  |  | (4.25) |
| $R^2$ | 0.45 | 0.45 |
| Observations | 1723 | 1723 |

Source: Based on Acemoglu and Johnson (2007)'s sample.

Note: OLS regressions with a full set of fixed effects. Robust standard errors, adjusted for clustering by country-disease pairs, are in parentheses.

Sample Diseases: Diphtheria, Influenza, Cholera, Typhoid, Smallpox, Shigella dysentery, Whooping cough, Measles, Scarlet fever, Yellow fever, Plague, and Typhus.

*p<.1

**p<.05

***p<.01

Table 2 reports the estimate of equation (4). We can conclude that the relationship between actual disease mortality and interventions is negative and statistically significant. In other words, the mortality from a certain disease declined with the intervention. In column 2, we add the previous interventions in regression; our result remains significant and consistent with our assumption. Additionally, our $R^2$ is 0.45 in both cases.

## C. First-Stage

We apply equation (3) to test our first-stage assumption for predicted mortality. Table 3 shows that there are negative and statistically significant correlations between our instrumental variable and Log life expectancy, whether in the base sample of 47 countries or poor countries. In column 2, the estimate of $\gamma$ is -0.31 (0.07), which means that an improvement of 0.1 in predicted mortality leads to a 3.1% increase in life expectancy. Comparing this with column 1, we conclude that the life expectancy of low- and middle-income countries somehow benefited less from international disease interventions.

**Table 3**
**First-Stage Estimates: Predicted Mortality and Log Life Expectancy**

|  | 1940 & 1980 | | |
| --- | --- | --- | --- |
|  | Base Sample | Low- & Middle-Income Countries | Base Sample: Interaction with Continent Dummies |
| Predicted mortality | -0.45*** | -0.31*** | -0.30*** |
|  | (0.06) | (0.08) | (0.07) |
| $R^2$ | 0.89 | 0.92 | 0.93 |
| Observations | 94 | 72 | 94 |
| Number of Countries | 47 | 36 | 47 |

Source: Based on Acemoglu and Johnson (2007)'s sample.

Note: OLS regressions with a full set of fixed effects. Robust standard errors, adjusted for clustering, are in parentheses.

*p<.1

**p<.05

***p<.01

## D. Main Results

Our main result from the replication is estimated using 2SLS and still focuses on the long-difference between 1940 and 1980.

Table 4 shows that the relationship between Log GDP and Log life expectancy is statistically insignificant both in our 47 countries base sample and in low- and middle-income countries.

We now interpret our main result from Panel B in table 4: we see a statistically significant and negative relationship between Log GDP per capita and Log life expectancy using predicted mortality as our instrument. In column 1, the estimate of $\alpha$ in equation (1) is -1.32 (0.56), showing that a 1% increase in life expectancy leads to a 1.32% decrease in GDP per capita. Somehow, in column 2, this negative relationship is amplified in low- and middle-income countries.

In replicating Acemoglu and Johnson (2007), we can conclude that there is no positive effect of health improvement on income. One rationale is that although health improvement may increase human capital and workers' productivity, lower mortality may increase the size of the population, thereby reducing other input factors per capita (Bloom, Canning and Fink, 2014, p. 1367-1368). In other words, the effect of life expectancy improvement on total GDP growth is not sufficient to compensate for the effect on increasing population.

**Table 4**
**Effect of Life Expectancy on GDP and per capita GDP: 2SLS Estimates**

|  | 1940 & 1980 | |
|---|---|---|
|  | Base Sample | Low- and Middle-Income Countries Only |
|  | A. Log GDP | |
| Log Life Expectancy | 0.32 | -0.39 |
|  | (0.84) | (1.44) |
| Number of Countries | 47 | 36 |
| $R^2$ | 0.97 | 0.96 |
|  | B. Log GDP per capita | |
| Log Life Expectancy | -1.32** | -2.35** |
|  | (0.56) | (1.13) |
| Number of Countries | 47 | 36 |
| $R^2$ | 0.92 | 0.87 |

Source: Based on Acemoglu and Johnson (2007)'s sample.

Note: 2SLS regressions that include a full set of fixed effects. Robust standard errors are in parentheses.

*p<.1

**p<.05

***p<.01

### E. Further Replications and Discussions

*i. Comments by Bloom, Canning and Fink (2014)*

Bloom, Canning and Fink (2014) comment on Acemoglu and Johnson (2007) using the argument that Acemoglu and Johnson did not introduce initial health status as a control variable in their regressions. Bloom, Canning and Fink (2014, p. 1360) believe that initial health status is significantly related to economic growth. Thus, in their differential model, they re-tested this problem with the addition of initial health.

We now use a difference form of equation (1) by adding a lagged independent variable in our model: initial life expectancy. Hence:

$$\Delta y_{it} = \varphi \Delta x_{it} + \delta_i + \omega x_{i,t-1} + \varepsilon_{it} \quad (5)$$

We estimate equation (5) with the change in predicted mortality and compare the results with those of Acemoglu and Johnson (2007).

Column 4 in table 5 shows that the instrumental variable from Acemoglu and Johnson (2007) failed to produce a statistically significant estimate with the addition of our initial health condition. However, column 3 estimates a simple OLS regression without an instrumental variable by adding initial health and shows that the correlation between growth in life expectancy and growth in GDP per capita somehow becomes positive and statistically significant.

**Table 5**
**Effect of Growth in Life Expectancy on Growth in GDP per capita**

|  | 1940-2000 | | | |
| --- | --- | --- | --- | --- |
|  | Acemoglu & Johnson | | Adding Initial Health | |
|  | OLS | IV | OLS | IV |
|  | (1) | (2) | (3) | (4) |
| Growth in life expectancy 1940-2000 | -1.14*** | -1.51*** | 3.68*** | -21.56 |
|  | (0.35) | (0.40) | (1.30) | (81.29) |
| Log Life Expectancy 1940 |  |  | 3.769*** | -15.23 |
|  |  |  | (0.94) | (61.37) |
| Constant | 1.71*** | 1.86*** | -14.89*** | 69.19 |
|  | (0.16) | (0.19) | (4.17) | (271.40) |
| Observations | 47 | 47 | 47 | 47 |
| $R^2$ | 0.19 | 0.17 | 0.33 | -3.12 |
| Cragg-Donald F-Statistics |  | 60.84 |  | 0.14 |
| Critical value for F-statistics |  | 16.38 |  | 16.38 |

Source: Based on the Acemoglu and Johnson (2007) sample.

Note: Robust standard errors are in parentheses. In specifications (2) and (4), our independent variable is instrumented with the change in predicted mortality.

*p<.1

**p<.05

***p<.01

Bloom, Canning and Fink (2014, p.1364) argue that countries with poor initial health have been experiencing slow economic growth since World War II. As countries with lower initial life expectancy experience the fastest health gains because new medical technologies can be used to address their high disease burden, the negative correlation between health gains and economic growth observed by Acemoglu and Johnson has emerged. If this is the case, we cannot provide a causal interpretation for the relationship between health improvement and income growth, as poor initial health leads to slower growth in income.

*ii. Reply by Acemoglu and Johnson (2014)*

Acemoglu and Johnson (2014) replied to Bloom, Canning and Fink (2014) by arguing that the introduction of initial health conditions in Bloom, Canning and Fink (2014) leads to a more severe multicollinearity bias.

To address this problem, we estimate equation (5) and add initial life expectancy from 1900, but include the interaction of time dummies and Lags of life expectancy from 1900 to 1940.

In column 1 of table 6, we use decadal data to estimate our equation (1) and find similar statistically significant results as those in table 4, although the magnitude of our estimate changes slightly. Column 2 shows the estimate by adding the interaction of Lags of life expectancy and time dummies; the magnitude and significance become smaller but remains negative. In addition, we include Lags of GDP per capita in our estimate in column 3 for comparison and the results are consistent with our expectation.

**Table 6**
**Effect of Life Expectancy on Log GDP per capita with controlling Initial Health: 2SLS**

|  | Dependent Variable: Log GDP per Capita 1940-2000 | | |
| --- | --- | --- | --- |
|  | Baseline Specification | Adding Lags of Life expectancy | Adding Lags of GDP per capita |
|  | (1) | (2) | (3) |
| Life expectancy | -1.394*** | -0.928* | -1.317** |
|  | (0.362) | (0.486) | (0.627) |
| Countries | 47 | 47 | 47 |
| Periods | 7 | 7 | 7 |
| $R^2$ | 0.77 | 0.80 | 0.80 |

Source: Based on the Acemoglu and Johnson (2007) sample.

Note: Robust standard errors are in parentheses.

*p<.1

**p<.05

***p<.01

## IV. Extension

### A. Bias from Population Size

In replicating Acemoglu and Johnson (2007)'s analysis, we used panel data of 47 countries at two time points (1940 and 1980) for a total of 94 observations. In my view, since we are analysing the relationship between health improvement and economic growth, we cannot ignore the considerable difference in population size between developed and underdeveloped countries. On the other hand, I believe that analysing the long-term impact of health improvement on economic growth in underdeveloped countries is more worthwhile. Thus, population size is incorporated in our new estimate, whereby we estimate equation (1) weighted by population size.

$$y_{it} = \alpha x_{it} + \eta_i + \mathbf{Z'_{it}\beta} + \mu_t + \varepsilon_{it} \qquad (1)$$

By population weighting, results show that the estimated coefficients were changed by a relatively small magnitude and are still statistically significant. Table 7 shows that our estimates in equation (1) with population weighting are different from the estimates in table 4 and that $R^2$ increases in all cases. In column 1, we can see that with population weighting the effect of Log life expectancy on Log per capita GDP becomes stronger and more statistically significant.

Although it is inaccurate to weight population directly, as we ignore the standard deviation of our variables within a country, we believe the bias from population size is larger and well worth eliminating.

**Table 7**
**Effect of Log Life Expectancy on Log per capita GDP: 2SLS (Population Weighted)**

|  | Base Sample | | Low- and Middle-Income Countries | |
| --- | --- | --- | --- | --- |
|  | 1940 & 1980 | 1940 & 2000 | 1940 & 1980 | 1940 & 2000 |
| Log Life expectancy | -1.659*** | -1.702*** | -2.171*** | -2.943*** |
|  | (0.001) | (0.001) | (0.002) | (0.001) |
| Number of Observations | 4755229 | 5843502 | 4003999 | 5014882 |
| R² | 0.983 | 0.968 | 0.967 | 0.963 |

Source: Based on the Acemoglu and Johnson (2007) sample. Note: Standard errors are in parentheses.

*p<.1

**p<.05

***p<.01

## B. Decadal Percentage Change IV Estimate

We now focus on the simple percentage change form estimate:

$$\frac{\Delta Y_{it}}{Y_{it}} = \frac{\varphi \Delta X_{it}}{X_{it}} + \sigma_i + \varepsilon_{it} \qquad (6)$$

where *X* is life expectancy and *Y* is GDP per capita. In this case, we use the difference form of predicted mortality (predicted mortality change) as our instrumental variable in this estimate.

We begin by roughly estimating the relationship between health improvement rate and economic growth rate between 1940 and 2000. Compared with column 2 in table 5, the result from column 1 in table 8 shows a significantly larger $R^2$. However, column 2 shows that the impact of health improvement on income growth in low- and middle-income countries is twice as large as that in our base sample. One rationale to interpret this result is that rich countries have a higher initial GDP per capita and life expectancy, and the same amount of change in GDP per capita leads to a significantly different change in GDP per capita growth rate between initially poor countries and rich countries.

**Table 8**
**Effect of Growth Rate of Life Expectancy on Growth Rate of per capita GDP: 2SLS**

|  | Base Sample | Low- and Middle-Income Countries |
|---|---|---|
|  | 1980 | 1980 |
|  | (1) | (2) |
| Life Expectancy Growth Rate | -3.313 *** | -6.813*** |
|  | (1.163) | (2.052) |
| Number of Observations | 47 | 36 |
| $R^2$ | 0.707 | 0.676 |
| Cragg-Donald F-Statistics | 42.151 | 12.165 |
| Critical value for F-statistics | 16.38 | |

Source: Based on the Acemoglu and Johnson (2007) sample. Note: Robust standard errors are in parentheses.

*p<.1

**p<.05

***p<.01

Next, we redo the above estimate using decadal data. Table 9 shows that the estimates of $\varphi$ are somehow smaller and still statistically significant. However, the $R^2$ declines significantly in both cases than in table 8. We will provide an explanation for this phenomenon in next section.

**Table 9**
**Effect of Growth Rate of Life Expectancy on Growth Rate of per capita GDP: 2SLS**

|  | Base Sample | Low- and Middle-Income Countries |
|---|---|---|
|  | 1940-1980 | 1940-1980 |
| Life Expectancy Growth Rate | -1.427*** | -1.504*** |
|  | (.466) | (.509) |
| Number of Observations | 342 | 276 |
| $R^2$ | 0.388 | 0.336 |
| Cragg-Donald F-Statistics | 77.670 | 57.843 |
| Critical value for F-statistics | 16.38 | |

Source: Based on the Acemoglu and Johnson (2007) sample.

Note: Robust standard errors are in parentheses.

*p<.1

**p<.05

***p<.01

In addition, we perform this process again with population weighting. Comparing the results presented in table 10 with those in table 8 and table 9, column 1 and column 3 show that the difference in the coefficient estimate of $\varphi$ between the main sample and developing countries decreases in the replicating estimate. The results presented in column 2 and column 4 support this finding when using a decadal data set in our 2SLS regressions.

More importantly, the difference of our estimate between the decadal specification and the 40-year specification is larger than that in table 9. Column 2 suggests that a 0.01 increase in the life expectancy growth rate decreases the per capita GDP growth rate by 0.06 and that this negative impact rises in developing countries, as column 4 shows. By comparing columns 1 and 3 with columns 2 and 4, we find that the effect of improved health on the personal income growth rate in the decadal estimate is 3 times as large as that in our replicating estimate. This unignorable difference implies that our original replication model omits the fluctuation in economic growth and the population differences across countries, that is, economic development is not stable and linear and developing countries should have larger weights in the regression.

In conclusion, the effect of improved health on income growth is actually larger when we regress our variables with decadal data using population size weighting. Developing countries benefit more from the global epidemiological transition, which may lead to a more significant negative correlation between our dependent and independent variables in the estimates.

**Table 10**
**IV 2SLS Estimate: Dependent Variable: Change Rate of per capita GDP**

|  | Main Sample | | Developing Countries | |
| --- | --- | --- | --- | --- |
|  | 1980 Only | 1940 to 1980 | 1980 Only | 1940 to 1980 |
| Life Expectancy Growth Rate | -2.130*** | -6.165*** | -2.623*** | -7.723*** |
|  | (.002) | (.0184) | (.004) | (.0319) |
| Number of Observations | 3140913 | 10362653 | 2692386 | 8792268 |
| $R^2$ | 0.697 | -10.186 | 0.637 | -18.962 |
| *C-D F-Statistics* | 4.9e+06 | 1.1e+05 | 3.1e+06 | 57.843 |

Source: Based on the Acemoglu and Johnson (2007) sample.

Note: Robust standard errors are in parentheses.

*p<.1

**p<.05

***p<.01

## C. Weak Instrument Test

In addition, we apply a weak instrument test by incorporating Cragg-Donald F-Statistics in table 8, table 9 and table 10 in the above extensions. We find that all the Cragg-Donald F-Statistics are significantly larger than the Stock-Yogo weak identification test critical value (16.38) with a 10% maximal IV size, except for specification (2) in table 8, which is larger than the critical value (8.96), with a 15% maximal IV size. Overall, we confirm that no matter whether we use our extended percentage change form or the replicated original construction, our IV is not weak in the first-stage.

## D. Review of IV

Acemoglu and Johnson (2007) constructed predicted mortality instrument based on international disease interventions; we now review their construction of this instrumental variable:

$$M_{it}^I = \sum_{d \in D} [(1 - I_{dt})M_{dit} + I_{dt}M_{dFt}] \qquad (2)$$

Equation (2) shows that our predicted mortality is actually the sum of the mortality rates from the intervened and non-intervened diseases at time t. Acemoglu and Johnson use zero as the mortality rate of all sample diseases after international intervention.

All 14 diseases they incorporated are infectious diseases. Table 11 shows the intervention dates for these base sample diseases. We can see that 12 of the diseases were curable with antibiotics. Specifically, with the introduction of penicillin, a huge proportion of nearly incurable infectious diseases became curable and a huge amount of lives were saved. This made our intervention rate for all observations 12/14 in 1940. Moreover, with the development of vaccines, the intervention rate reached 100% in 1960.

We review the intervention dummy variable that Acemoglu and Johnson (2007) constructed ($I_{dt}$ is 1 after the international intervention) in our zeroth-stage data. Although it is not exactly consistent with our main regression data, there is no difference after we take zero as the value of mortality from a certain disease with an intervention in the construction of predicted mortality.

There is a serious problem, as shown in table 12 and figure 3; when we use this dummy variable in our first-stage test of instrumental variable, we find that the predicted mortality becomes the sum of these 14 initial mortalities without international interventions after accounting for our difference form estimate in equation (5). In other words, predicted mortality is zero after 1950 for all cases.

As all 14 diseases are infectious, both the initial GDP per capita and economic growth rate of a country must be related to the initial mortality from all 14 diseases. Countries with a higher average income have better medical conditions and sanitation facilities, so people are less likely to become infected. In contrast, bad health condition has a negative impact on economic growth. Thus, we should conclude that the exclusion restriction of our instrumental variable is violated.

**Table 11**
**Summary of Interventions**

| Disease | Treatment | Intervention Time |
|---|---|---|
| Tuberculosis | Antibiotics | 1940 |
| Pneumonia | Antibiotics | 1940 |
| Influenza | Antibiotics | 1940 |
| Cholera | Antibiotics | 1940 |
| Typhoid | Antibiotics | 1940 |
| Smallpox | Vaccination | 1950 |
| Shigella dysentery | Antibiotics | 1940 |
| Whooping cough | DTP Vaccination and Antibiotics | 1940 |
| Measles | Vaccination | 1940 |
| Diphtheria | Antibiotics | 1940 |
| Scarlet fever | Antibiotics | 1940 |
| Yellow fever | Vaccination | 1950 |
| Plague | Antibiotics | 1940 |
| Typhus | Antibiotics | 1940 |

Source: Based on the Acemoglu and Johnson (2007) sample.

**Table 12**
**Predicted Mortality in Base Sample**

| Year | Observations | Mean | Standard Deviation | Minimum Value | Maximum Value |
|---|---|---|---|---|---|
| 1940 | 91 | .435 | .284 | .030 | 1.198 |
| 1950 | 91 | .005 | .015 | 0 | .101 |
| 1960 | 91 | 0 | 0 | 0 | 0 |
| 1970 | 91 | 0 | 0 | 0 | 0 |
| 1980 | 91 | 0 | 0 | 0 | 0 |
| 1990 | 91 | 0 | 0 | 0 | 0 |
| 2000 | 91 | 0 | 0 | 0 | 0 |
| Total | 637 | .063 | .186 | 0 | 1.198 |

Source: Based on the Acemoglu and Johnson (2007) sample.

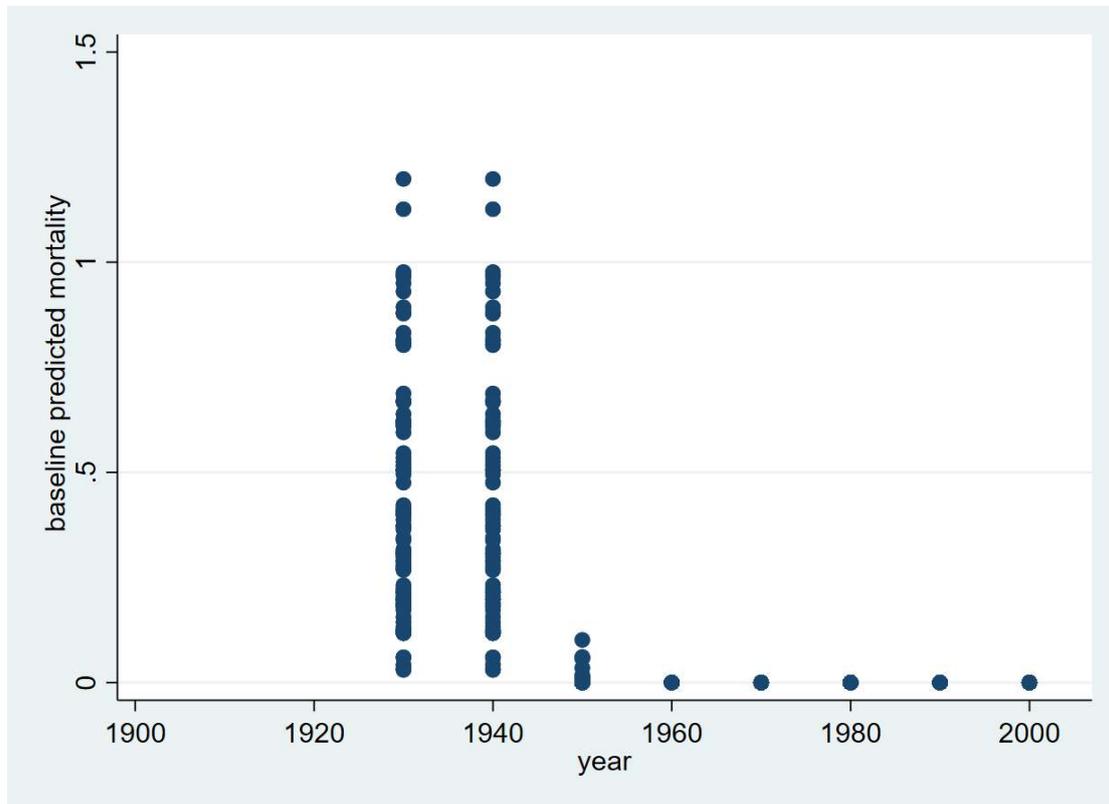

Fig. 3. Scatter Plot of Predicted Mortality

Source: Based on Acemoglu and Johnson (2007) sample.

To interpret this mathematically, we rewrite equation (2) for every sample year:
In 1940, the value of our instrumental variable is equal to:

$$M^{I}_{i1940} = \sum_{d \in D} M_{di1940}$$

In 1950, the value of our instrumental variable is equal to (***s*** is Smallpox; ***y*** is Yellow fever):

$$M^I_{i1950} = M_{si1950} + M_{yi1950}$$

And after 1950, predicted mortality is zero in all sample years:

$$M^I_{i196} = M^I_{i1970} = M^I_{i1980} = M^I_{i1990} = M^I_{i2000} = 0$$

As we only estimate the change between 1940 and 1980 or 1940 and 2000, we can reconfirm that our instrumental variable is actually the initial mortality from all 14 diseases without international disease interventions.

This concern can be shown using a simple OLS estimate without any other control variables in tables 13 and 14. Table 13 shows that the effect of income on initial total mortality from these diseases is significantly negative with an $R^2$ of approximately 0.47 as our expectation; specifically, a 1% increase in GDP per capita is associated with a 0.6% improvement of initial total mortality. In table 14, we can conclude that the effect of initial mortality from the sample diseases is significantly negative on GDP per capita growth rate. Clearly, many other factors affect income growth rate, but we can conclude that higher initial total mortality slows per capita GDP growth using our simple OLS estimate. Although logically we cannot statistically test the validity of exclusion restriction for our instrumental variable, tables 13 and 14 provide evidence consistent with our expectation.

**Table 13**
**Effect of Log GDP per capita on Log Pre-intervened Mortality**

|  | Base Sample |
|---|---|
| Log GDP per capita | -.627*** |
|  | (.103) |
| Constant | 3.904*** |
|  | (.806) |
| $R^2$ | 0.473 |
| Number of Observations | 47 |

Source: Based on the Acemoglu and Johnson (2007) sample.

Note: Robust standard errors are in parentheses.

*p<.1

**p<.05

***p<.01

**Table 14**
**Effect of Pre-intervened Mortality on GDP per capita growth rate between 1940 and 1950**

|  | Base Sample |
|---|---|
| Pre-intervened Mortality | -.252** |
|  | (.123) |
| Constant | .231*** |
|  | (.070) |
| $R^2$ | 0.08 |
| Number of Observations | 56 |

Source: Based on the Acemoglu and Johnson (2007) sample.

Note: Robust standard errors are in parentheses.

*p<.1

**p<.05

***p<.01

We now estimate our first-stage regression using decadal panel data once again. The reason our coefficients became larger and larger in absolute value in table 15 from columns 1 to 3 can be interpreted by the scatter plot of our first stage test (Figure 4). That is, adding more zero-valued time-series dependent variables (predicted mortality) can change the slope of our OLS estimate. The longer our estimating period, the more zero values of predicted mortality added, the steeper our estimating line. Although such estimates are statistically significant in our constrained sample, it is economically meaningless; for example, we obtain a significantly smaller coefficient if we only regress for 1940 and 1950, as column 1 shows. We can now interpret the results in table 9; that is, in our decadal percentage change estimate (Eq. (6)), the value of our instrumental variable (the decadal change in predicted mortality) becomes zero after 1950.

**Table 15**
**First-Stage Estimate: Effect of Predicted Mortality on Life Expectancy**

|  | 1940-1950 | 1940-1980 | 1940-2000 |
|---|---|---|---|
| Predicted Mortality | -0.18*** | -0.33*** | -0.41*** |
|  | (0.05) | (0.06) | (0.06) |
| $R^2$ | 0.72 | 0.81 | 0.83 |
| Number of Observations | 106 | 283 | 401 |
| Number of Countries | 59 | 59 | 59 |

Source: Based on the Acemoglu and Johnson (2007) sample.

Note: Robust standard errors are in parentheses.

*p<.1

**p<.05

***p<.01

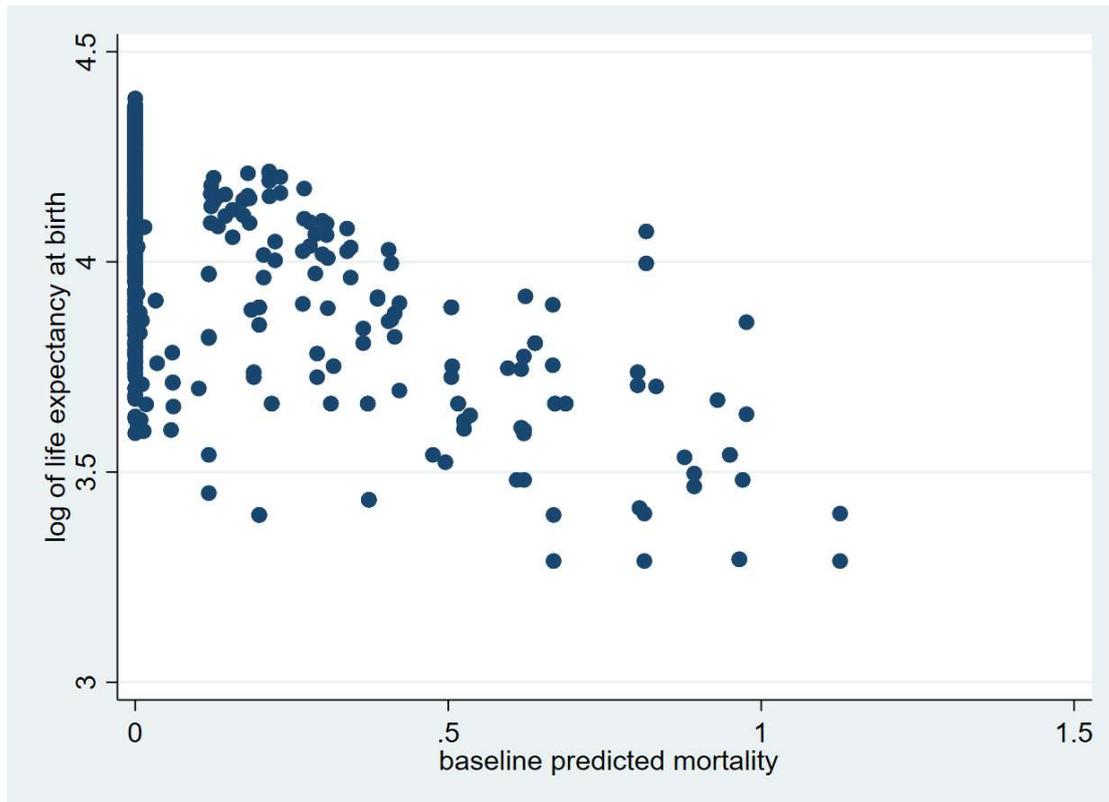

Fig. 4. Y: Log life expectancy; X: Predicted Mortality

Source: Based on Acemoglu and Johnson (2007) sample.

Even without our interpretation of the endogeneity of predicted mortality in our main regression, we can still refute their zeroth-stage in macro views. There are three reasons why the WHO international disease interventions claimed by Acemoglu and Johnson (2007) should have no relationship in the construction of our instrumental variable (Eq. (2)) with our dependent variable (GDP per capita or per capita GDP growth rate):

First, innovations of medical science. Although technology improvement is uncertain, it does have some unfunctionally strong relationships with GDP growth. Intervention for different diseases was not carried out simultaneously. The capacity and efficiency of transferring medical science and technology to medical production is highly related to the GDP growth of OECD countries (especially the U.S.). Furthermore, as mentioned earlier, 12 of 14 diseases were curable by penicillin; the price of one unit of penicillin was more than 200 dollars in 1943 but declined to less than 0.5 dollars in 1947. If we take the ratio of the cost of such medical services and GDP per capita, we can consider Acemoglu and Johnson (2007)'s first claim invalid, and although the change in medical cost played a major role in this ratio, we believe that certain medical costs are generally related to productivity growth. On the other hand, this innovation in the medical industry, with its use of antibiotics and vaccinations, will increase the proportion of medical industry in the total GDP. (Figure A1 in the Appendix shows the change in price and production of penicillin between 1945 and 1975; Figure A2 shows the change in current health expenditure (CHE) as a percentage of GDP in recent years and Figure A3 roughly reports the change in the total cost of medical services between 1935 and 1964 in the U.S.)

Second, the establishment of the World Health Organization (WHO). The constitution of the WHO was signed in 1946 and became involved in Egyptian cholera the following year. In the early decades after the WHO's establishment, the WHO's program budget is financed mainly through the assessed contributions that are the dues countries pay to be a member of the organization; the amount each member state must pay is calculated relative to the country's wealth and population (the WHO website). Furthermore, people come together usually to obtain greater profits, and we cannot ignore the fact that all these intervened sample diseases are worldwide infectious. Given the source of WHO's budget, OECD countries played a significant role during the period of international disease interventions. That is, annual budget distribution was related to our dependent variable. Furthermore, as the WHO carried out this huge project over many years, we cannot assume that the costs they incurred and the transfer of medical-care staffs throughout various countries had no impact on the local and global economies.

Third, the change in international values. Acemoglu and Johnson (2007) claimed that somehow people or countries become more willing to help others. However, if we could not feed ourselves, we would not commit suicide to help others. Wealthy countries become rich first, then help less developed countries. More importantly, as mentioned above, all 14 sample diseases are infectious, one reason countries help others is to avoid being affected by these diseases, as these could possibly do enormous damage to a country's economy. Consequently, the change in international value is associated with economic growth.

More essentially, based on my statistical interpretation, the effect of our dummy variable (international disease interventions) on zeroth-stage disappears when we estimating short- or long-difference specifications; that is, international disease interventions have no relationship with our instrumental variable.

*E. Further Discussion*

At present, many countries suffer from population aging. However, all the above estimates omit the difference in the current demographic structure across countries. Acemoglu and Johnson (2007) demonstrate a positive effect of life expectancy on total births, but they do not consider the population size. We now estimate the effect of life expectancy growth on fertility rate.

**Table 16**
**Effect of Life Expectancy on Fertility Rate, 1940-2000**

|  | Base Sample | Population Weighting |
|---|---|---|
| Log Life Expectancy | -.0471*** | -.0534*** |
|  | (.00147) | (0.00001) |
| Number of Observations | 574 | 11910075 |
| $R^2$ | 0.643 | 0.764 |

Source: Based on the Acemoglu and Johnson (2007) sample.

Note: Robust standard errors are in parentheses.

*p<.1

**p<.05

***p<.01

The negative correlation between fertility rate and life expectancy in table 16 suggests a possible phenomenon: the healthier people are, the fewer babies they want to give birth. If this is the case, persistently low fertility rates in countries with relatively longer life expectancy will aggravate the population aging, thus reducing the productivities per capita in the long-run. We will do further research to address this issue by adding the current demographic structure as the control variable in our estimate.

## V. Conclusion

Our primary aim is to explain the invalidity of the instrumental variable in Acemoglu and Johnson (2007), not to invent a new mechanism showing how health affects income. Evidently, their instrumental variable violates the principle of exclusion restriction, which leads to a greater bias in our estimate. Our results show that such invalid instrumental variables counterproductively exaggerate the influence of independent variables on dependent variables. This exaggerated effect of life expectancy on economic growth generates a more severe reverse causality bias in estimates. In other words, the initial mortality in a low- or middle-income country has a greater impact on our dependent variable (economic growth), which is why the greater negative correlation between life expectancy and income emerges in low- and middle-income countries in our estimates.

## References

Acemoglu, D and Johnson, S 2007, 'Disease and Development: The Effect of Life Expectancy on Economic Growth.', *Journal of Political Economy,* vol. 115, no. 6, pp. 925–85.

Acemoglu, D and Johnson, S 2014, 'A reply to Bloom, Canning and Fink', *Journal of Political Economy,* vol. 122, no. 6, pp. 1367-1375.

Bloom, D, Canning, D and Fink, G 2014, 'Disease and Development Revisited', *Journal of Political Economy,* vol. 122, no. 6, pp. 1355-1366.

Gallup, JL and Sachs, JD 2001, 'The Economic Burden of Malaria.', *American J. Tropical Medicine and Hygiene,* vol. 64, suppl. 1, pp. 85–96.

Roser, M 2015, 'Life Expectancy', *OurWorldInData.org website*, Retrieved 1 October, 2019 from: *http://ourworldindata.org/data/population-growth-vital-statistics/life-expectancy/*

Preston, SH 1975, 'The Changing Relation between Mortality and Level of Economic Development.', *Population Studies* vol. 29, no. 2, pp. 231–48.

Klepper, S and Simons, KL 1997, 'Technological Extinctions of Industrial Firms: An Inquiry into their Nature and Causes.', *Industrial and Corporate Change*, vol. 6, no. 2, pp. 379-460.

Barzel, Y 1969, 'Productivity and the Price of Medical Services', *Journal of Political Economy,* vol. 77, no. 6, pp. 1014-1027.

The World Health Organization, Retrieved 1 October, 2019 from: *https://www.who.int/about/finances-accountability/funding/assessed-contributions/en/*

# Appendix

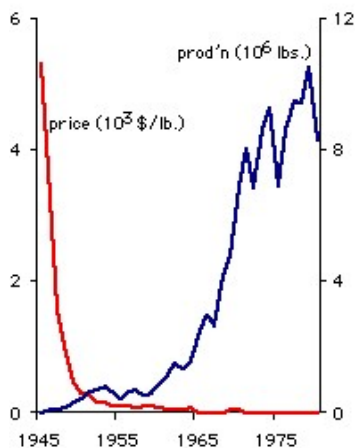

Figure A1 Source. – **Klepper. S and Simons. KL (1997, pp. 379-460).**

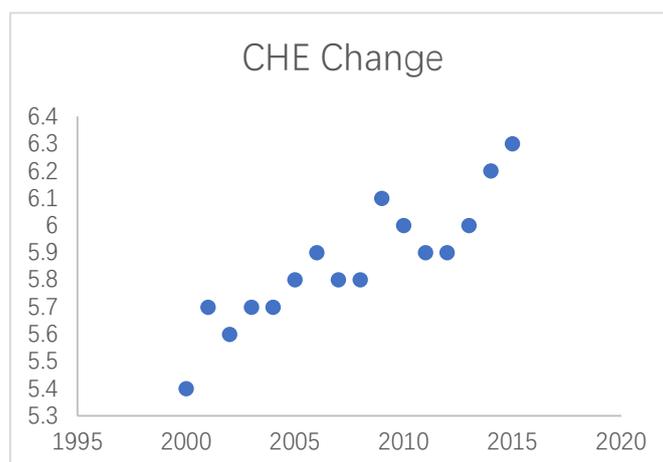

Figure A2 Source. – **Based on the WHO.**

THE CPI AND SOME OF ITS MEDICAL COMPONENTS

| Time Period | All Items | All Medical Care | Physicians' Fees | Hospital Daily Service Charges |
|---|---|---|---|---|
| 1935–64 (1935 = 110) | 226.2 | 241.17 | 217.6 | 608.8 |
| 1945–64 (1945 = 100) | 172.4 | 207.7 | 185.3 | 445.8 |

SOURCE.—U.S. Department of Labor, Bureau of Labor Statistics (relevant issues).

Figure A3 Source. – **Barzel. Y (1969, p. 1014).**